\documentclass[twocolumn,prl,amsmath,amssymb,floatfix,superscriptaddress,showpacs,longbibliography]{revtex4-1}
\usepackage{graphicx}
\usepackage{dcolumn}
\usepackage{bm}
\usepackage{array}
\usepackage{booktabs}
\usepackage[flushleft]{threeparttable}
\usepackage{xspace}
\usepackage{hyperref}
\usepackage{color}
\def\new{\color{black}}

\def\rucl{$\alpha$-RuCl$_3$\xspace}
\begin{document}

\title{{\color{black}Evidence for} magnetic fractional excitations in a Kitaev quantum-spin-liquid candidate $\alpha$-RuCl$_3$}
\author{Kejing~Ran}
\altaffiliation{These authors contributed equally to the work.}
\author{Jinghui~Wang}
\altaffiliation{These authors contributed equally to the work.}
\affiliation{School of Physical Science and Technology and ShanghaiTech Laboratory for Topological Physics, ShanghaiTech University, Shanghai 200031, China}
\affiliation{National Laboratory of Solid State Microstructures and Department of Physics, Nanjing University, Nanjing 210093, China}
\author{Song~Bao}
\author{Zhengwei~Cai}
\author{Yanyan~Shangguan}
\affiliation{National Laboratory of Solid State Microstructures and Department of Physics, Nanjing University, Nanjing 210093, China}
\author{Zhen~Ma}
\affiliation{National Laboratory of Solid State Microstructures and Department of Physics, Nanjing University, Nanjing 210093, China}
\affiliation{Institute for Advanced Materials, Hubei Normal University, Huangshi 435002, China}
\author{Wei~Wang}
\affiliation{School of Science, Nanjing University of Posts and Telecommunications, Nanjing 210023, China}
\author{Zhao-Yang~Dong}
\affiliation{Department of Applied Physics, Nanjing University of Science and Technology, Nanjing 210094, China}
\author{P.~\v{C}erm\'{a}k}
\affiliation{J\"{u}lich Centre for Neutron Science (JCNS) at Heinz
Maier-Leibnitz Zentrum (MLZ), Forschungszentrum J\"{u}lich GmbH,
Lichtenbergstr. 1, 85748 Garching, Germany}
\affiliation{Charles University, Faculty of Mathematics and Physics, Department of Condensed Matter Physics, Ke Karlovu 5, 121 16, Praha, Czech Republic}
\author{A.~Schneidewind}
\affiliation{J\"{u}lich Centre for Neutron Science (JCNS) at Heinz
Maier-Leibnitz Zentrum (MLZ), Forschungszentrum J\"{u}lich GmbH,
Lichtenbergstr. 1, 85748 Garching, Germany}
\author{Siqin~Meng}
\affiliation{Helmholtz-Zentrum Berlin f\"{u}r Materialien und Energie GmbH, Hahn-Meitner-Platz 1D-14109 Berlin, Germany}
\affiliation{China Institute of Atomic Energy, Beijing 102413, China}
\author{Zhilun~Lu}
\affiliation{Helmholtz-Zentrum Berlin f\"{u}r Materialien und Energie GmbH, Hahn-Meitner-Platz 1D-14109 Berlin, Germany}
\affiliation{The Henry Royce Institute and Department of Materials Science and Engineering, The University of Sheffield, Sir Robert Hadfield Building, Sheffield, S1 3JD, United Kingdom}
\author{Shun-Li~Yu}
\email{slyu@nju.edu.cn}
\author{Jian-Xin~Li}
\email{jxli@nju.edu.cn}
\author{Jinsheng~Wen}
\email{jwen@nju.edu.cn}
\affiliation{National Laboratory of Solid State Microstructures and Department of Physics, Nanjing University, Nanjing 210093, China}
\affiliation{Collaborative Innovation Center of Advanced Microstructures, Nanjing University, Nanjing 210093, China}


\begin{abstract}
$\alpha$-RuCl$_3$ has been studied extensively because of its proximity to the Kitaev quantum-spin-liquid (QSL) phase and the possibility of approaching it by tuning the competing interactions. Here we present the first polarized inelastic neutron scattering study on \rucl single crystals to explore the scattering continuum around the $\Gamma$ point at the Brillouin zone center, which was hypothesized to be resulting from the Kitaev QSL state but without concrete evidence. With polarization analyses, we find that  while the spin-wave excitations around the M point vanish above the transition temperature $T_{\rm N}$, the pure magnetic continuous excitations around the $\Gamma$ point are robust against temperature. Furthermore, by calculating the dynamical spin-spin correlation function using the cluster perturbation theory, we derive magnetic dispersion spectra based on the $K$-$\Gamma$ model, which involves with a ferromagnetic Kitaev interaction of $-7.2$~meV and an off-diagonal interaction of $5.6$~meV. We find this model can reproduce not only the spin-wave excitation spectra around the M point, but also the non-spin-wave continuous magnetic excitations around the $\Gamma$ point. {\new These results provide evidence for the existence of fractional excitations around the $\Gamma$ point originating from the Kitaev QSL state, and further support the validity of the $K$-$\Gamma$ model as the effective minimal spin model to describe $\alpha$-RuCl$_3$.}

\end{abstract}


\maketitle
Quantum spin liquids (QSLs) with fractionalized spin excitations and long-range entanglement have drawn a lot of attention since Anderson first proposed the QSL state within the resonating-valence-bond (RVB) model~\cite{Anderson1973153} and used it to explain the high-temperature superconductivity~\cite{anderson1}. Typically, QSL states have been proposed in triangular and kagome lattices where antiferromagnetic interactions are highly frustrated due to the geometrical constrain, and therefore there remain strong quantum fluctuations in these systems that prevent spins from establishing a long-range order as observed in a usual magnet at low temperatures~\cite{nature464_199,RevModPhys.89.025003,0034-4885-80-1-016502,npjqm4_12,Broholmeaay0668}. Besides the RVB model which was built upon geometrical frustration, the Kitaev model defined on the ideal two-dimensional honeycomb lattice with spin $S$= 1/2, is an exactly solvable model with the QSL ground state~\cite{aop321_2}. On a honeycomb lattice, there is no geometrical frustration, and it is the anisotropic bond-dependent Kitaev interactions that lead to the frustration on a single site and give rise to the ``Kitaev" QSL state~\cite{aop321_2,arXiv:1701.07056}. A Kitaev QSL can host fractional excitations represented by {\it e.g.}, Majorana fermions, which behave as anyons obeying the non-Abelien
statistics under magnetic field, and allow to be braided for fault-tolerant topological quantum
computation~\cite{Kitaev20032,aop321_2,RevModPhys.80.1083,Barkeshli722}.

In general, for a spin-only system, it is unrealistic to find the bond-dependent Kitaev interactions which underlie the Kitaev QSL, as the spin interactions ought to be isotropic along the three symmetry-equivalent bonds in a honeycomb lattice. One plausible approach of realizing the anisotropic Kitaev interaction is to resort to spin-orbital-coupling (SOC) assisted Mott insulators, where the effective moment $J_{\rm eff}=1/2$ is an entanglement of the orbital moment $L=1$ and spin moment $S=1/2$~\cite{prl102_017205,arcmp7_195,arXiv:1701.07056}. This idea was initially applied to iridates such as Na$_2$IrO$_3$ where iridium and
oxygen ions form edge-shared IrO$_6$ octahedra, and the 5$d$ iridium ions with strong SOC and $J_{\rm eff}=1/2$ form a two-dimensional honeycomb network~\cite{prl102_017205,prl105_027204,PhysRevB.84.180407,PhysRevLett.108.127203,prl110_097204,PhysRevLett.112.077204,PhysRevB.90.155126,np11_462,0953-8984-29-49-493002}. The geometrical configuration suppresses the isotropic Heisenberg interaction while on the other hand promotes the anisotropic Kitaev interaction in the aid of the large SOC and strong spatial anisotropy of the $d$ orbitals~\cite{prl102_017205,PhysRevB.93.214431}. These features should have made iridates a promising platform to investigate the Kitaev physics. However, these materials suffer from the following problems: i) the presence of monoclinic and trigonal
distortions renders the applicability of the localized $J_{\rm eff}$ picture to
these materials questionable~\cite{PhysRevLett.109.197201,PhysRevB.88.035107,0953-8984-29-49-493002}; ii) iridium has a large neutron absorption cross section and thus it is difficult to carry out inelastic neutron scattering (INS) measurements to extract the magnetic interactions; iii) large-size and high-quality single crystals are not available~\cite{0953-8984-29-49-493002}.

\begin{figure}[ht]
\centering
\includegraphics[width=0.95\linewidth,trim=0mm 0mm 0mm 0mm]{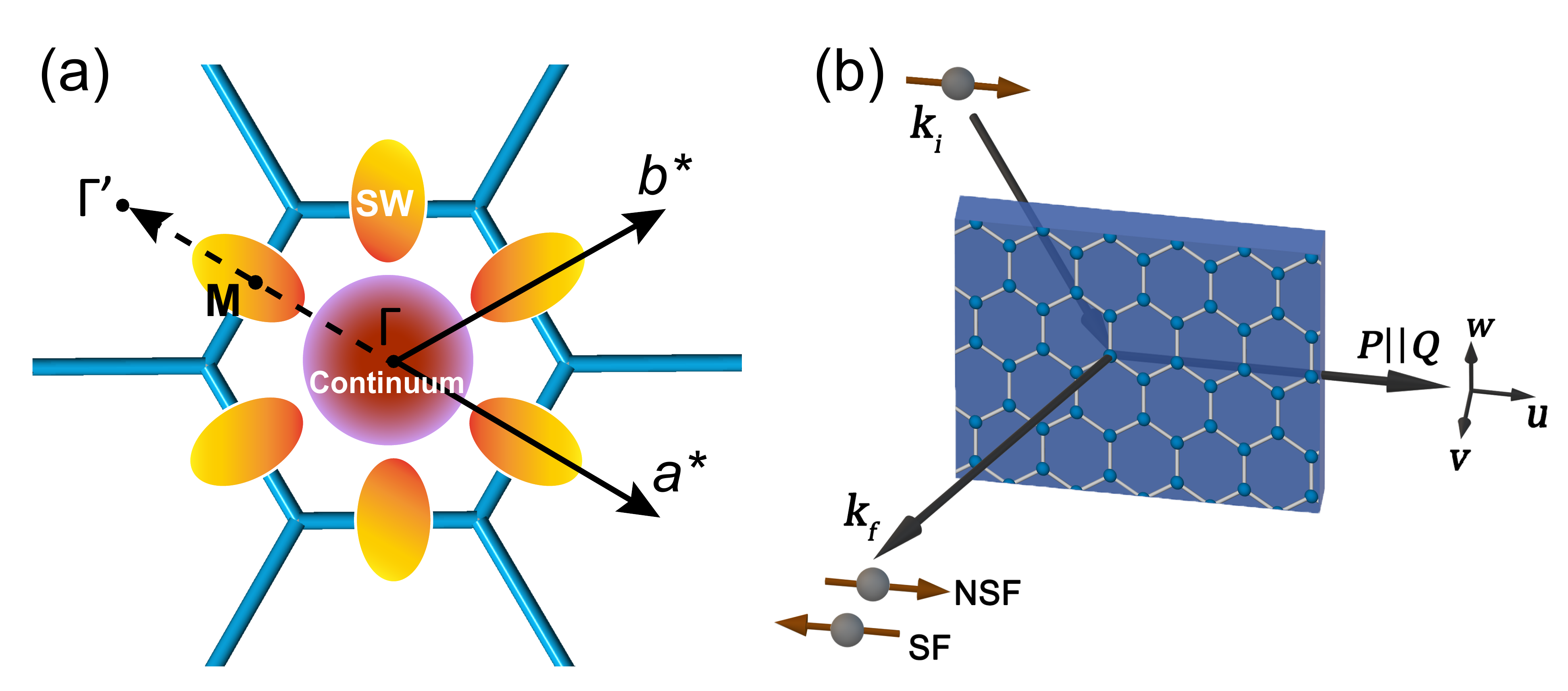}
\caption{\label{fig1}{(a) Schematic of the magnetic excitations in the first Brillouin zone for \rucl in the zigzag magnetic order state. Thick solid lines denote the Brillouin zone boundary. Excitations around the $\Gamma$ point represent the scattering continuum. The six-fold symmetric excitations around the M point denote spin-wave excitations. (b) Schematic of the experimental setup for the polarized neutron scattering measurements for the case of $\bm{P}\|\bm{Q}$. $\bm{P}$ is the neutron polarization direction, and $\bm{Q}$ is the scattering vector, obtained by $\bm{Q}=\bm{k}_{\rm{i}}-\bm{k}_{\rm{f}}$. $\bm{k}_{\rm{i}}$ and $\bm{k}_{\rm{f}}$ are the initial and final wave vector of neutrons, respectively. Neutron polarization directions parallel, up and perpendicular, with respect to $\bm{Q}$ are labeled by $\bm{u}$, $\bm{w}$ and $\bm{v}$, respectively. SF and NSF denote spin-flip and non-spin-flip of neutrons, respectively.}} 
\end{figure}

Therefore, more efforts have been devoted into \rucl with the honeycomb lattice recently, which can also realize the SOC-assisted $J_{\rm eff}=1/2$ Mott insulating state, despite the weaker SOC effect of the 4$d$ Ru$^{3+}$~\cite{PhysRevB.90.041112,PhysRevB.91.144420}. Although the low-temperature phase is not the long-sought Kitaev QSL but zigzag ordered phase instead~\cite{PhysRevB.91.144420,PhysRevB.92.235119,nm15_733,PhysRevB.93.134423,PhysRevLett.118.107203}, it is shown to exhibit salient Kitaev interactions due to the close-to-ideal bond configuration in the RuCl$_6$ octohedron~\cite{PhysRevB.91.241110,PhysRevB.93.075144,PhysRevB.93.155143,nm15_733,PhysRevLett.118.107203,PhysRevB.96.115103,PhysRevB.96.064430,Banerjee1055,PhysRevB.97.075126,np16_837}. Notably, the zigzag order is close to the Kitaev QSL phase in the phase diagram and is fragile subject to external magnetic field~\cite{PhysRevB.91.094422,PhysRevB.92.235119,PhysRevB.91.180401,PhysRevB.95.180411,PhysRevB.95.245104,PhysRevLett.120.067202,npjqm3_8,PhysRevB.100.060405} or pressure~\cite{PhysRevB.96.205147,PhysRevB.97.241108,PhysRevB.97.245149}. More intriguingly, a moderate in-plane magnetic field may drive the system into a QSL phase~\cite{PhysRevLett.119.037201,PhysRevLett.119.227208,PhysRevLett.120.187201,nature559_227}. In the zigzag phase, INS experiments show that the excitations are mostly concentrated around the M and $\Gamma$ points of the two-dimensional Brillouin zone, as sketched in Fig.~\ref{fig1}(a). While it is generally believed the gapped sharp excitations around the M point are spin waves associated with the magnetic order, the origin of the broad continuum around the $\Gamma$ point is still controversial~\cite{nc8_1152,PhysRevLett.114.147201,np12_912,PhysRevLett.119.227201,PhysRevLett.119.227202,PhysRevB.96.165120,nm15_733,Banerjee1055,np13_1079,PhysRevLett.118.107203}. 
{\color{black}On the one hand, various spectroscopic results, such as Raman scattering~\cite{PhysRevLett.114.147201,np12_912}, optical THz spectroscopy~\cite{PhysRevLett.119.227201,PhysRevLett.119.227202,PhysRevB.96.165120}, and INS spectra~\cite{nm15_733,Banerjee1055,np13_1079}, show a nearly temperature-independent excitation continuum around the $\Gamma$ point, which is reminiscent of the fractionalized excitations originating from the Kitaev QSL state. On the other hand, because of the small scattering angle around the $\Gamma$ point, the broad continuous feature in the INS experiment could be due to the nuclear scattering~\cite{Banerjee1055}, which is also temperature independent. In addition, it has been argued that the continuum feature may originate from the incoherent excitations caused by strong magnetic anharmonicity~\cite{nc8_1152}. Therefore, it is critical to distinguish whether the continuum is fractional magnetic excitation or nuclear scattering.}
Polarized neutron scattering experiment from which pure magnetic scattering can be obtained will help solve this problem~\cite{p1959,PhysRev.181.920}. However, because of the much reduced scattering cross section in a polarized neutron experiment, it is very challenging to obtain meaningful results for thin single crystals like \rucl.

\begin{figure}[ht]
\centering
\includegraphics[width=0.9\linewidth,trim=15mm 0mm 15mm 1mm]{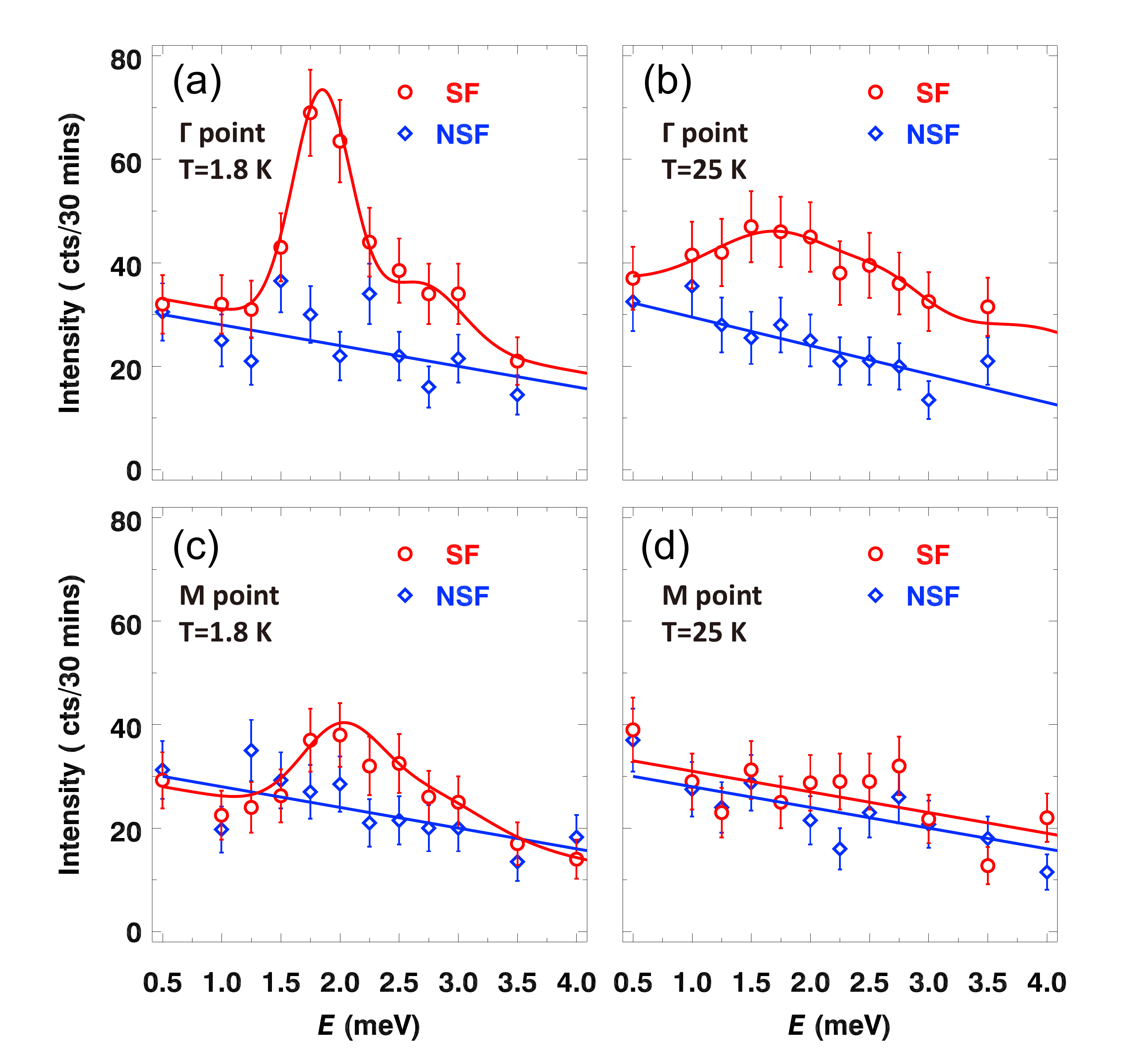}
\caption{\label{fig2}{(a) and (b) Energy scans of \rucl at the $\Gamma$ point (0,\,0,\,1.5) measured at 1.8 and 25~K with polarized neutrons in the spin-flip (SF) and non-spin-flip (NSF) modes. The neutron polarization used was $\bm{P}\|\bm{Q}$. (c) and (d) Same scans as in (a) and (b) but at the M point (0.5,\,0,\,0). Lines through data are guides to the eye. Errors represent one standard deviation throughout the paper.}}
\end{figure}

\begin{figure}[ht]
\centering
\includegraphics[width=0.95\linewidth,trim=0mm 0mm 0mm 1mm]{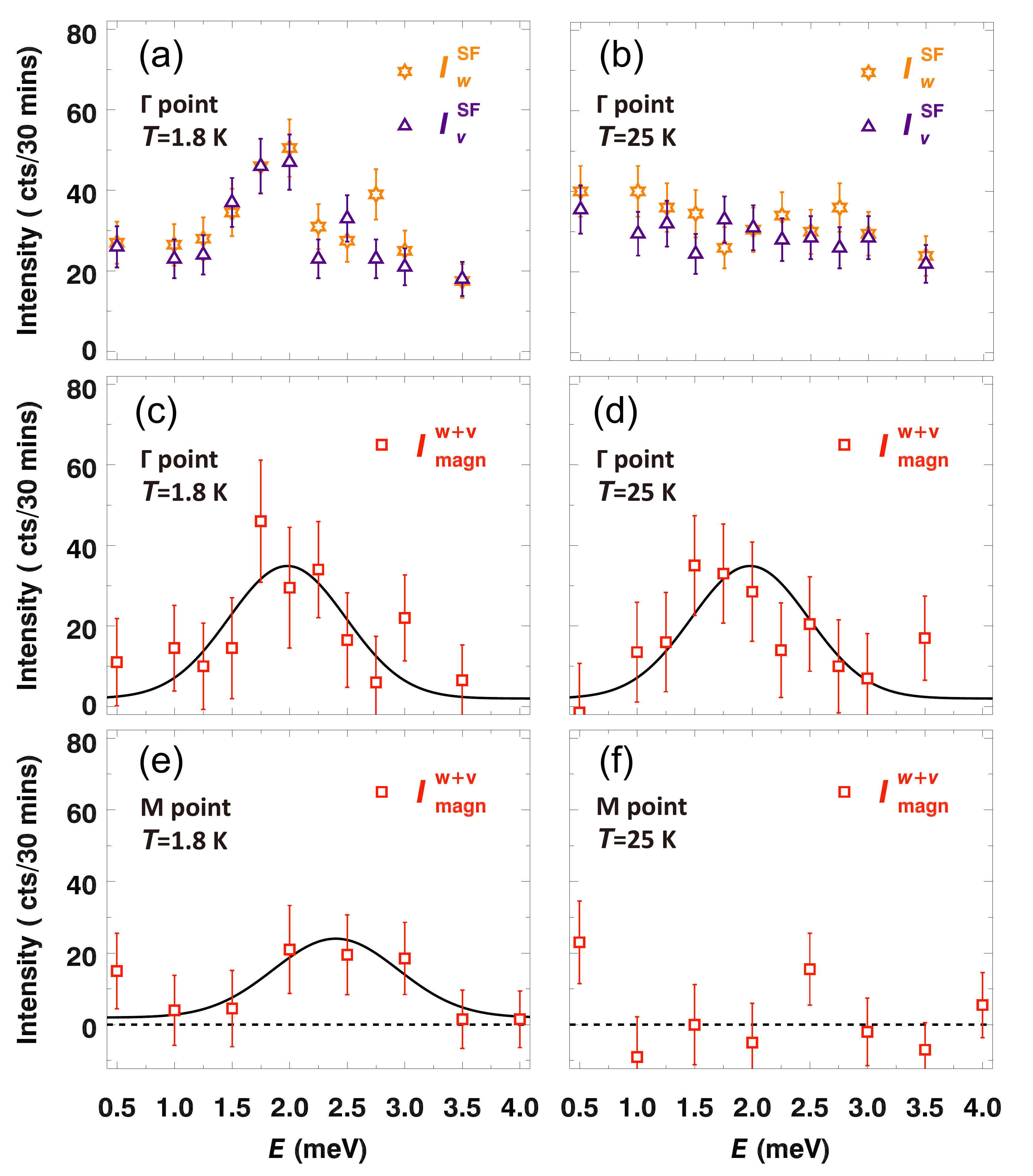}
\caption{\label{fig3}{(a) and (b) Energy scans at the $\Gamma$ point measured at 1.8 and 25~K with polarized neutrons in the SF mode. The neutron polarization used was $\bm{P}\perp\bm{Q}$. ${I}_{v}^{\rm{SF}}$ and ${I}_{w}^{\rm{SF}}$ are scattering intensities for SF mode in directions of perpendicular and up with respect to $\bm{Q}$. (c) and (d) Pure magnetic intensities around the $\Gamma$ point measured at 1.8 and 25~K, obtained by ${I}_{\rm{mag}}^{w+v}={I}_{\rm{mag}}^{w}+{I}_{\rm{mag}}^{v}=2{I}_{u}^{\rm{SF}}-{I}_{v}^{\rm{SF}}-{I}_{w}^{\rm{SF}}$. (e) and (f) Same as in (c) and (d) but at the M point. Solid lines through data are guides to the eye. Dashed lines denote the background.}}
\end{figure}

\begin{figure*}
  \centering
  \includegraphics[width=0.95\linewidth]{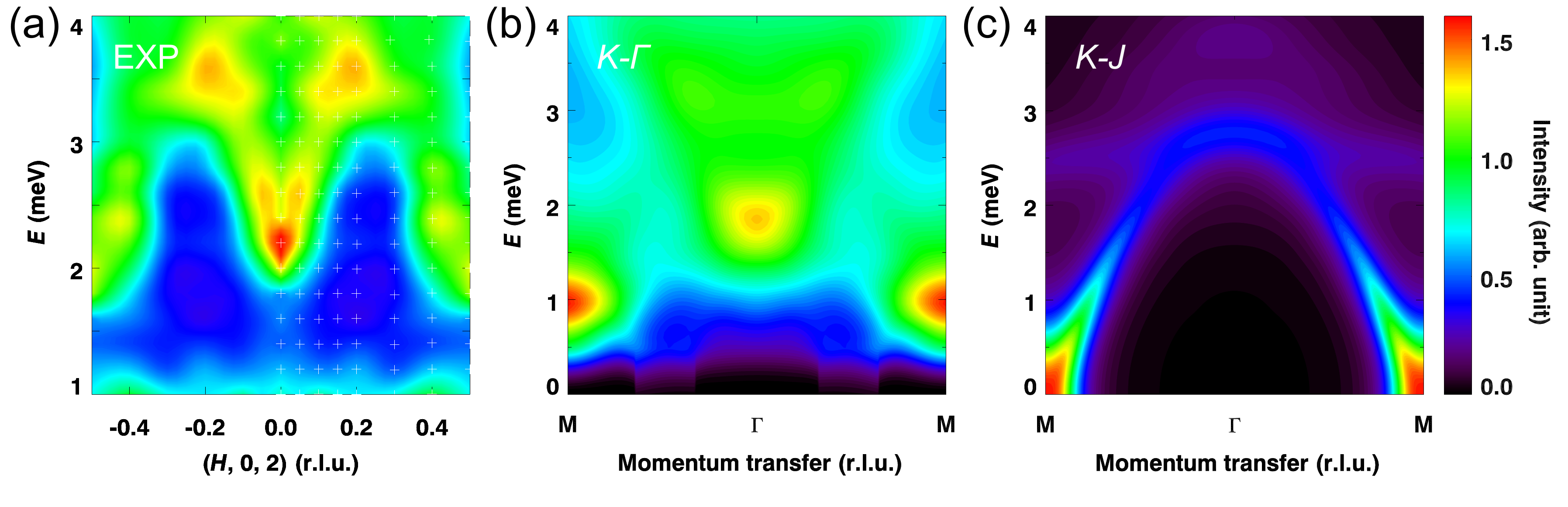}\\
  \caption{\label{theory}{(a) Magnetic dispersion along [$H$00] at T=1.8 K obtained by plotting the points marked with white cross dots. Left side of the dispersion from $H$=-0.5 to 0 r.l.u is the mirror symmetrical figure on the right from $H$=0 to 0.5 r.l.u. (b) Magnetic excitation spectrum of the $K$-$\Gamma$ model for $\Gamma/K=-0.73$ with $K<0$. (c) Theoretical magnetic excitation spectra for the zigzag phases of the $K$-$J$ model for $J/K=-0.36$ with $K>0$.}}
\end{figure*}

In this Letter, we report polarized neutron scattering study on 2-g high-quality \rucl single crystals. The data allow us to distinguish magnetic scattering from nuclear scattering. Our polarization results show that the broad continuous excitations around the $\Gamma$ point are of magnetic origin, which are persistent at temperatures far above the magnetic ordering temperature $T_{\rm{N}}$. In contrast, the spin-wave excitations around the M point disappear above $T_{\rm{N}}$. Moreover, by comparing the calculation results using Kitaev-off-diagonal ($K$-$\Gamma$) and Kitaev-Heisenberg ($K$-$J$) models with the unpolarized neutron experimental data, we find that the $K$-$\Gamma$ model with a ferromegntic $K=-7.2$~meV and $\Gamma=5.6$~meV can describe both the low-energy response of the spin-wave excitations associated with the zigzag ordered state around the M point, and the continuous magnetic excitations around the $\Gamma$ point. These results are consistent with the presence of fractional magnetic excitations in \rucl and further support the $K$-$\Gamma$ model to be the minimal effective spin model in describing the system.

High-quality single crystals of \rucl were grown using the chemical-vapor-transport method with \rucl powders. The sheet-like crystals had a natural $a$-$b$ plane as the cleavage plane and had a typical size of $10\times10\times1$~mm$^3$, which weighed about 60~mg a piece. Both susceptibility measurements and specific heat measurements showed a sharp antiferromagnetic phase transition at $\sim$7.5 K~\cite{PhysRevLett.118.107203,PhysRevLett.119.227208,PhysRevLett.120.067202}. Polarized neutron scattering measurements were performed on PANDA, a cold neutron triple-axis spectrometer with the polarized option located at FRM II~\cite{panda}. $XYZ$-difference method for neutron polarization analysis~\cite{Schaerpf} was used to separate the magnetic scattering from nuclear coherent, nuclear spin incoherent, isotopic incoherent and background contributions. Unpolarized neutron scattering measurements were carried out on a cold neutron triple-axis spectrometer FLEXX located at HZB. We used a fixed final energy mode with $E_{\rm f}=5$~meV, and a double-focusing condition for both the monochromator and analyzer for all these experiments. For the experiment on PANDA, about 45 single crystals weighed about 2~g in total were co-aligned together on two aluminum plates using a Laue x-ray diffractometer. Heusler alloys with the (1,\,1,\,1) reflection plane were used as the monochromator and analyzer. Guide magnetic fields were used to rotate the neutron polarization direction along ($\bm{P}\|\bm{Q}$) or vertical to the scattering wave vector ($\bm{P}\perp\bm{Q}$). A Mezei coil flipper was used between the sample and analyzer to flip the polarization of neutrons. The flipping ratios of the three directions $\bm{u}$, $\bm{v}$ and $\bm{w}$ we labeled in Fig.~\ref{fig1}(b) were 18.1, 7.5, 16.9, respectively. We denote a scattering event in which the spin makes a 180$^\circ$ rotation as spin flip (SF), and scattering with no spin direction change as non spin flip (NSF). For the measurements on FLEXX, we used 24 single crystals with a total mass of $\sim$2.2~g. All the experiments were conducted in the $(H0L)$ plane. {\color{black}The sample mosaic was about 3.6 degrees at PANDA and 1.93 degrees at FLEXX scanned around the Bragg peak (0,\,0,\,3).} We used the P3$_1$12 notation with $a=b=5.96$~\AA, and $c=17.2$~{\AA} as the lattice constants. The wave vector \textbf{\emph{Q}} was described by ($HKL$) reciprocal lattice units (r.l.u.) of $(a^{*}, b^{*}, c^{*})=(4\pi/\sqrt3a, 4\pi/\sqrt3b, 2\pi/c)$.


According to the polarized neutron scattering rules~\cite{neutron1}, when the neutron polarization direction $\bm{P}$ is parallel to the scattering vector $\bm{Q}$, magnetic and nuclear scatterings are detected in the SF and NSF channels,
respectively. In the experiment, we polarized neutrons along the scattering vector first. Here we take (0,\,0,\,1.5) as the $\Gamma$ point since neutron scattering experiment cannot access the $\Gamma$ point at $\bm{Q}=0$. Figure~\ref{fig2} shows energy scans at the $\Gamma$ and M points (0.5,\,0,\,0) at 1.8~K (below the $T_{\rm{N}}$ of 7.5~K) and 25~K (well above the $T_{\rm{N}}$). As shown in Fig.~\ref{fig2}(a), pronounced magnetic excitations are observed in the SF channel at $T=1.8$~K at the $\Gamma$ point. There is a clear peak centered at about 1.8~meV. On the other hand, the data in the NSF channel is essentially a sloping background. These results indicate that the excitations at the $\Gamma$ point detected in the unpolarized INS experiments previously are of magnetic origin~\cite{Banerjee1055,np13_1079,npjqm3_8}. When we increase the temperature up to 25~K, in the SF channel, there are still clear magnetic excitations on top of the sloping background, as shown in Fig.~\ref{fig2}(b). The magnetic excitations are significantly broader than those in the magnetically ordered state. This broad and continuum-like feature is a characteristic of the fractional excitations~\cite{Banerjee1055,np13_1079}. {\new We suspect that the reason why intensities at low temperature are stronger than those at high temperature is that the low-temperature excitations have some interactions with the spin-wave excitations.} For comparison, we also measured the excitations at the M point, which is the wave vector of the zigzag magnetic order~\cite{PhysRevB.91.144420,PhysRevLett.118.107203}. As shown in Fig.~\ref{fig2}(c), a magnetic peak is observed at $T=1.8$~K in the SF channel on top of the sloping background obtained in the NSF channel. As expected, this peak disappears when the temperature is above the $T_{\rm{N}}$, consistent with it being the spin-wave excitations arising from the zigzag magnetic order~\cite{nm15_733,PhysRevLett.118.107203,Banerjee1055}. The different temperature dependence of the excitations at the $\Gamma$ and M points and the persistence of the magnetic excitations at the $\Gamma$ point further prove that these exciations are magnetic, fractionalized, and robust within a certain temperature window~\cite{np13_1079,PhysRevResearch.2.043015}.

Although all the magnetic scattering is present in the SF channel when $\bm{P}$ $\|$ $\bm{Q}$, the total intensity contains some background resulting from the incoherent nuclear spin scattering with an intensity contribution of $\frac{2}{3}{{I}_{\rm{inc}}^{\rm{spin}}}$~\cite{neutron1}. In order to deduct this term and obtain pure magnetic scattering intensities, we polarized the neutron beam to the two directions $\bm{v}$ and $\bm{w}$ perpendicular to $\bm{Q}$ as illustrated in Fig.~\ref{fig1}(b) and performed the same scans as in Fig.~\ref{fig2} to obtain ${I}_{v}^{\rm{SF}}$ and ${I}_{w}^{\rm{SF}}$, as shown in Fig.~\ref{fig3}(a) and (b). Pure magnetic scattering intensities are given by ${I}_{\rm{mag}}^{w+v}={I}_{\rm{mag}}^{w}+{I}_{\rm{mag}}^{v}=2{I}_{u}^{\rm{SF}}-{I}_{v}^{\rm{SF}}-{I}_{w}^{\rm{SF}}$~\cite{neutron1}. The so-obtained intensities at the $\Gamma$ point are shown in Fig.~\ref{fig3}(c) and (d). It can be seen that they exist at both low and high temperatures. The temperature-stable magnetic excitations at the $\Gamma$ point are evident of the fractional excitations which are originated from the Kitaev QSL state. For comparison, the pure magnetic intensities at the M point are shown in Fig.~\ref{fig3}(e) and (f). In contrast to those at the $\Gamma$ point, the magnetic excitations at the M point are only observed at the low temperature. {\new The observations that the pure magnetic intensities ${I}_{\rm{mag}}^{w+v}$ at the $\Gamma$ point remain constant against temperature while those of spin waves at the M point are completely vanishing above the $T_{\rm N}$ indicate the emergence of fractional excitations at a higher temperature above the $T_{\rm N}$~\cite{np13_1079,PhysRevResearch.2.043015}.}


After pinning down the origin of the excitations at the $\Gamma$ point, we next discuss the effective spin model to describe the magnetic excitations at both the $\Gamma$ and M points, in contrast to previous attempts where the excitations around the two momenta were treated separately using different models~\cite{PhysRevLett.118.107203,PhysRevB.96.115103,nm15_733,Banerjee1055}. To do this, we first performed unpolarized INS measurements at $T=1.8$~K to obtain the magnetic dispersion as shown in Fig.~\ref{theory}(a). There are clear magnetic excitations weakly dispersing from both the $\Gamma$ and M points. A broad continuum with a gap of $\sim$1.8~meV appears around the $\Gamma$ point, consistent with the polarization data and earlier results~\cite{Banerjee1055,np13_1079,npjqm3_8}. To describe these data, we first consider the $K$-$\Gamma$ model that we used to fit the spin-wave excitations in our previous works~\cite{PhysRevLett.118.107203,PhysRevB.96.115103}. Its Hamiltonian can be written as,
\begin{equation}
H_{K-\Gamma}=\sum_{<ij>}[KS^{\gamma}_{i}S^{\gamma}_{j}+\Gamma(S^{\alpha}_{i}S^{\beta}_{j}+S^{\beta}_{i}S^{\alpha}_{j})].
\label{model-kg}
\end{equation}
Here, $\gamma=x,y,z$ labels the three nearest-neighbor bonds on the honeycomb lattice, and $(\alpha,\,\beta,\,\gamma)$ is a permutation of $(x,y,z)$ with $\alpha$ and $\beta$ labeling the two remaining directions on a $\gamma$ bond. We calculated the dynamical spin-spin correlation function using the cluster perturbation theory, which has been successfully applied to other magnetically frustrated systems~\cite{PhysRevB.98.134410}.
The calculated results with one typical parameter set of ferromagnetic $K=-7.2$~meV and $\Gamma=5.6$~meV are shown in Fig.~\ref{theory}(b). These parameters ensures that the system is in the zigzag order phase~\cite{PhysRevLett.112.077204,PhysRevB.96.115103}, and are consistent with our previous works~\cite{PhysRevLett.118.107203,PhysRevB.96.115103}. Compared to the experimental spectra shown in Fig.~\ref{theory}(a), it is clear that both the excitations at the M and $\Gamma$ points can be well captured by this model. {\new As we demonstrated in Refs.~\cite{PhysRevLett.118.107203,PhysRevB.96.115103}, the spin-wave excitations should only be observed to disperse from the M point around 2~meV and reach the band top at the $\Gamma$ point at $E\sim$4~meV within the $K$-$\Gamma$ model. In Fig.~\ref{theory}(a) and (b), the continuous excitation spectra around the $\Gamma$ point drop to a much lower energy of $\sim$2~meV, therefore they should not be coherent spin waves but represent fractional excitations, supporting our polarization analyses.}

For comparison, we also consider the widely proposed $K$-$J$ model for Kitaev materials~\cite{prl102_017205,prl105_027204,prl110_097204,PhysRevLett.112.077204}:
\begin{equation}
H_{K-J}=\sum_{<ij>}(J\bm{S}_{i}\cdot\bm{S}_{j}+KS^{\gamma}_{i}S^{\gamma}_{j}).
\label{model-kh}
\end{equation}
The calculated results with $K=9.8$~meV and $J=-3.6$~meV  using the cluster perturbation theory are shown in Fig.~\ref{theory}(c). Although the parameter set places the system in the correct zigzag order phase in the phase diagram constructed using the $K$-$J$ model~\cite{prl102_017205,prl105_027204,prl110_097204,PhysRevLett.112.077204}, and produces the spin-wave excitations dispersing from the M point, the calculation deviates from experimental observations in various aspects. First, the spin-wave excitations are gapless, in contrast to the gapped spectra shown in Fig.~\ref{theory}(a). Second, the low-energy excitations near the $\Gamma$ point are absent. Third, the Kitaev interaction is antiferromagnetic, contradicting previous reports on the ferromagnetic Kitaev interaction in \rucl~\cite{PhysRevLett.118.107203,PhysRevB.96.115103,Banerjee1055,np16_837}. Therefore, the comparison between the experimental spectra and the theoretical results from different models not only confirms that the $K$-$\Gamma$ model is a more appropriate minimal spin model to describe \rucl~\cite{PhysRevLett.118.107203,PhysRevB.96.115103,Banerjee1055}, but also represents as evidence that there present continuous fractional magnetic excitations near the $\Gamma$ point. We note, however, in addition to the dominant $K$ and $\Gamma$ terms, some other  finite longer-range terms, although small, may be necessary to stabilize the zigzag order state~\cite{PhysRevB.96.064430,PhysRevResearch.2.043015}.

To conclude, we study the magnetic excitations by carrying out polarized INS measurements on high-quality \rucl single crystals. We find that there exist pure continuous magnetic excitations near the $\Gamma$ point, which are robust against temperature, as opposed to the spin waves near the M point that vanish above $T_{\rm N}$. In addition, our calculation using the cluster perturbation theory shows that the $K$-$\Gamma$ model with a ferromegntic $K=-7.2$~meV and a comparable $\Gamma=5.6$~meV not only can reproduce the spin-wave excitations near the M point, but also the continuous excitations near the $\Gamma$ point. These results are evident that there exist exotic fractional excitations in \rucl due to its proximity to the Kitaev QSL state.

{\new However, considering the weak cross sections of the polarized experiment, further investigations with better statistics by using more single crystals, increasing the counting time, or optimizing the polarization options, should be helpful to solidify the conclusion. Potentially, one can learn details about the anisotropic interactions and magnetizations with more serious polarization analysis because the scattering intensities contain both an interference and a Chiral magnetic scattering term~\cite{Schweika_2010}. In Ref.~\onlinecite{PhysRevResearch.2.043015}, it was proposed that the bond-dependent fractional excitations in $\alpha$-RuCl$_3$ can be manifested in the equal-time structure factor $S^{\gamma}({\bm Q})$ ($\gamma=x,y,z$ represents different spin components) in the intermediate temperature range. We believe this is an interesting proposal worth of future efforts. Furthermore, for systems like \rucl, strongly spin-orbital-coupling may give rise to complex spin excitations like hybrid modes, the existence of frustrated anisotropic interactions will naturally lead to noticeable anharmonic effects~\cite{nc8_1152}. These may better explain the enhancement of the low-temperature excitations at the $\Gamma$ point as observed experimentally.}

The work was supported by National Key Projects for Research and Development of China with Grant No.~2021YFA1400400, the National Natural Science Foundation of China with Grant Nos.~11822405, 12074174, 12074175, 11774152, 11904170, 12004249, 12004251 and 12004191, Natural Science Foundation of Jiangsu Province with Grant Nos.~BK20180006, BK20190436 and BK20200738, Shanghai Sailing Program with Grant Nos.~ 20YF1430600 and 21YF1429200, Fundamental Research Funds for the Central Universities with Grant No.~020414380183, and the Office of International Cooperation and Exchanges of Nanjing University. The experiment at FLEXX was carried out under proposal No.~172-05993CR using beamtime allocated in the HZB-CIAE collaboration on the scientific use of instruments.

%

\end{document}